\documentclass[12p]{article}
\usepackage{eso-pic} 
\usepackage{latexsym}
\usepackage{color}
\usepackage{amsmath}
\usepackage{epsfig}
\usepackage{hyperref}
 \usepackage{dcolumn}
   \usepackage{threeparttable}

\newcommand{\ortala}[1]{\begin{center}#1\end{center}}

\newcommand{\integ}[3]{{{\underset{#1 }{\overset{#2}{\displaystyle\int}}}#3}}

\newcommand{\summ}[3]{{{\underset{#1 }{\overset{#2}{\displaystyle\sum}}}#3}}

\newcommand{\re}[1]{(\ref{#1})}

\newcommand{\eq}[2]{\begin{equation}\label{#1}  #2\end{equation}}

\newcommand{\paran}[1]{\left(#1\right)}

\newcommand{\sch}[1]{Schrodinger}

\newcommand{\komb}[2]{\paran{\begin{array}{c} #1 \\ #2 \end{array}}}

\linethickness{0.6pt}

\begin{document}

\ortala{\textbf{Effects of the randomly distributed magnetic field on the phase diagrams of the Ising Nanowire II: continuous distributions}}

\ortala{\textbf{\"Umit Ak\i nc\i \footnote{umit.akinci@deu.edu.tr}}}

\ortala{\textit{Department of Physics, Dokuz Eyl\"ul University,
TR-35160 Izmir, Turkey}}

\section{Abstract}

The effect of the random magnetic field distribution on the phase
diagrams and ground state magnetizations of the Ising nanowire has been investigated with effective field theory with correlations. Gaussian
distribution has been chosen as a random magnetic field distribution. The
variation of the phase diagrams with that distribution parameters has been
obtained and some interesting results have been found such as disappearance of the reentrant
behavior and first order transitions which appear in  the case of discrete distributions. Also for single and double Gaussian distributions , ground state
magnetizations for different distribution parameters have been determined
which can be regarded as separate partially ordered phases of the
system. Keywords: \textbf{Ising Nanowire; random
magnetic field; Gaussian magnetic field distribution}

\section{Introduction}\label{introduction}

Recently there has been growing interest both theoretically and
experimentally in the magnetic nanomaterials such as nanoparticles,
nanorods, nanotubes and nanowires. Nowadays, fabrication of these
nanomaterials is no longer difficult, since development of the
experimental techniques permits us making materials with a few atoms.
For instance, acicular magnetic nano elements were already fabricated
\cite{ref1,ref2,ref3} and  magnetization of the nanomaterial has
been measured \cite{ref4}. Nanoparticle systems have growing application areas, e.g. they can be
used as sensors \cite{ref60},
permanent magnets \cite{ref62}, beside some medical applications
\cite{ref61}. In particular, magnetic nanowires and nanotubes have
many applications in nanotechnology \cite{ref53,ref54}. Nanowires
can be used as an ultrahigh density magnetic recording media
\cite{ref36,ref37,ref59}   and they have potential applications in
biotechnology \cite{ref42,ref43}, such as Ni nanowires can be used
for bio seperation \cite{ref69,ref70}.

In the nanometer scale, physical properties of these finite materials are
different from those of their bulk counterparts. Some properties of these materials, which highly depend on the size and the
dimensionality, can be used for fabrication of materials for various purposes. From this point of view, it is important to determine the properties of these materials theoretically. Most common used theoretical methods
for determining the magnetic properties of these materials are
mean field approximation (MFA), effective field theory (EFT) and
Monte Carlo (MC) simulation as in bulk systems. For
instance,  nanoparticles investigated by EFT with correlations
\cite{ref44}, MFA and MC \cite{ref45}. The phase diagrams and the
magnetizations of the nanoparticle described by the transverse Ising
model have been investigated by using MFA and EFT \cite{ref34,ref35}. Moreover, investigation of compensation temperature of the nanoparticle \cite{ref67} and magnetic properties of the nanocube with MC \cite{ref33} are among these studies.

Another method, namely variational cumulant expansion (VCE) based on
expanding the free energy in terms of the action up to $m^{th}$ order,
has been applied to the magnetic superlattices \cite{ref10} and
ferromagnetic nanoparticles \cite{ref11,ref12}. The first order
expansion within this method gives the results of the MFA.

Various nano structures can be modeled by core-shell models and these models
can be solved also by MFA, EFT and MC such as $FePt$ and $Fe_3 O_4$ nanotubes
\cite{ref51}. The phase diagrams and
magnetizations of the transverse Ising nanowire has been treated within
MFA and EFT \cite{ref38,ref39}, the effect of the surface dilution
on the magnetic properties of the cylindrical Ising nanowire and
nanotube has been studied \cite{ref40,ref58}, the magnetic
propeties of nanotubes of different diameters, using armchair or
zigzag edges has been investigated with MC \cite{ref41}, initial
susceptibility of the Ising nanotube and nanowire have been calculated within the EFT
with correlations \cite{ref46,ref47} and the compensation
temperature which appears for negative core-shell coupling has been
investigated by EFT for nanowire and nanotube \cite{ref52}. There
are also some works dealing with hysteresis characteristics of the
cylindrical Ising nanowire \cite{ref57,ref71}. Beside these, higher
spin nanowire or nanotube systems have also been investigated, such as spin-1
nanotube \cite{ref63} and nanowire \cite{ref64}, mixed spin - $3/2,1$ core
shell structured nanoparticle \cite{ref65}, mixed spin - $1/2,1$
nanotube \cite{ref66} systems.

On the other hand, as far as we know, there have less attention paid on
quenched randomness effects on these systems, except the site
dilution. However, including quenched randomness or disorder effects
in these systems may induce some beneficial results. For this
purpose we investigate the effects of the random magnetic field
distributions on the phase diagrams of the Ising nanowire within
this work. As stated in \cite{ref40} the phase diagrams of the
nanotube and nanowire are qualitatively similar, then investigation
of the effect of the random magnetic field distribution on the
nanowire will give hints about the effect of the same distribution
on the phase diagrams of the nanotube.

The Ising model in a quenched random field (RFIM) has been studied
over three decades. The model which is actually based on the local
fields acting on the lattice sites which are taken to be random
according to a given probability distribution was introduced for the
first time by Larkin \cite{refs1} for superconductors and later
generalized by Imry and Ma \cite{refs2}.   Beside the similarities
between diluted antiferromagnets in a homogenous magnetic field and ferromagnetic systems in the presence of random
fields \cite{refs3,refs4}, the importance of the random
field distributions on these systems comes from  the fact that,
random distribution of the magnetic field drastically affects the phase
diagrams of the system, and hence the magnetic properties. This situation has been
investigated widely in the literature for the bulk Ising systems.
For example, using a Gaussian probability distribution, Schneider
and Pytte \cite{refbulk1}
 have shown that phase diagrams of the
model exhibit only second order phase transition properties.
On the other hand,
Aharony \cite{refbulk2} and Mattis \cite{refbulk3} have introduced bimodal and
trimodal distributions, respectively, and they have reported the
observation of tricritical behavior. With the same distributions and using EFT with correlations,
Borges and Silva \cite{refbulk4,refbulk5,refbulk6} showed that three dimensional lattices show tricritical behavior while two dimensional lattices do not exhibit this behavior. On the other hand, by using two site EFT instead of one site EFT, tricritical behavior can be observed on a  square lattice \cite{refbulk7}. Similarly, Sarmento and Kaneyoshi \cite{refbulk8}
investigated the phase diagrams of RFIM by means of EFT
with correlations for a bimodal field distribution, and they
concluded that reentrant behavior of second order is possible
for a system with ($q \ge 6$). Recently, Fytas et al. \cite{refbulk16} applied
 MC simulations on a simple cubic lattice. They found that the transition
is continuous for a bimodal field distribution, while
 Hadjiagapiou \cite{refbulk20}  observed reentrant behavior and confirmed
the existence of a tricritical point for an asymmetric bimodal
probability distribution within the MFA
based on a Landau expansion.

In a recent series of papers,
phase transition properties of infinite dimensional RFIM with
 symmetric double \cite{refbulk9} and triple \cite{refbulk10} Gaussian random
fields have also been studied by means of a replica method
and a rich variety of phase diagrams have been presented. The situation has also been handled on 3D lattices with
nearest-neighbor interactions by a variety of theoretical works
such as EFT \cite{refbulk11,refbulk12}, EFT with multi site spin correlations \cite{refbulk13}, MC simulations \cite{refbulk14,refbulk15,refbulk17}, pair approximation \cite{refbulk18}, and
the series expansion method\cite{refbulk19}.

As seen in the short literature in bulk Ising systems, random field distributions keep up to date in the literature. Thus the aim of this work is
to inspect the effects of random field distributions on the phase diagrams of the nanowire system as a nanostructure.
The paper is organized as follows: In Sec. \ref{formulation} we
briefly present the model and  formulation. The results and
discussions are presented in Sec. \ref{results}, and finally Sec.
\ref{conclusion} contains our conclusions.

\section{Model and Formulation}\label{formulation}

\begin{figure}[h]\begin{center}
\epsfig{file=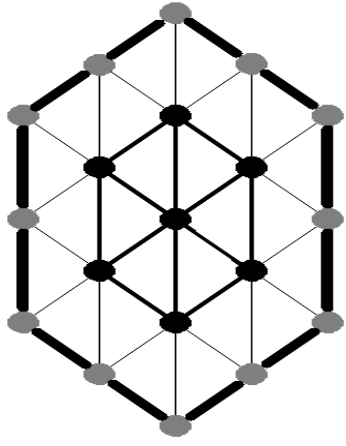, width=5cm,height=6cm}
\end{center}
\caption{Schematic representation of a cylindrical nanowire (top view). The gray/black circles represent the surface/shell magnetic atoms, respectively.}
\label{sek1}\end{figure}

We consider a nanowire which has geometry shown in Fig. \ref{sek1}.
The Hamiltonian of the nanowire is given by

\eq{denk1}{\mathcal{H}=-J_{1}\summ{<i,j>}{}{s_is_j}-J_{2}\summ{<m,n>}{}{s_ms_n}-J_3\summ{<i,m>}{}{s_is_m}-\summ{i}{}{H_is_i}-\summ{m}{}{H_ms_m}}
where $s_i$ is the $z$ component of the spin at a lattice site $i$ and it takes the values $s_i=\pm 1$ for the spin-1/2 system. $J_1$ and $J_2$ are the exchange
interactions between spins which are located at the core and shell,
respectively, and $J_3$ is the exchange interaction between the core
and shell spins which are nearest neighbor to each other. $H_i$ and
$H_m$ are the external longitudinal magnetic fields at the lattice
sites $i$ and $m$ respectively. Magnetic fields are distributed on the
lattice sites according to a given probability distribution. The first three
summations in Eq. \re{denk1} are over the nearest-neighbor pairs of
spins, and the other summations are over all the lattice sites.

This work -as a continuation of the earlier work \cite{refs8}- deals
with the following continuous magnetic field distribution,
\eq{denk2}{P\paran{H_i}=pG\paran{0,\sigma}+\frac{1-p}{2}\left[G\paran{H_0,\sigma}+G\paran{-H_0,\sigma}\right]
}where $G\paran{H_0,\sigma}$ is the Gaussian distribution centered at $H_{0}$ with a width $\sigma$ and  it is given by \eq{denk3}{
G\paran{H_0,\sigma}=\paran{\frac{1}{2\pi\sigma^2}}^{1/2}\exp{\left[-\frac{\paran{H_i-H_0}^2}{2\sigma^2}\right]}.
} Distribution given in Eq. \re{denk2} reduces to the system with zero magnetic field (pure system)
for $p=1,\sigma=0$. According to the distribution given in Eq.
\re{denk2}, $p$ percentage of the lattice sites are subjected to a
magnetic field chosen from the Gaussian distribution which has
$\sigma$ width and $H_0=0$ as a center. Half of the remaining sites
are under the influence of a field $H_i$  which is randomly chosen
from  the magnetic field distribution $G\paran{H_0,\sigma}$, whereas the distribution
$G\paran{-H_0,\sigma}$ used as distribution function on the
remaining sites.

Four different representative magnetizations ($m_i,i=1,2,3,4$) for the system can be given by usual EFT equations which are obtained by differential operator technique and decoupling approximation (DA) \cite{refs5,refs6},
\eq{denk4}{\begin{array}{lcl}
m_1&=&\left[A_1+m_1B_1\right]^4\left[A_3+m_2B_3\right]\left[A_3+m_3B_3\right]^2\left[A_1+m_4B_1\right]\\
m_2&=&\left[A_3+m_1B_3\right]\left[A_2+m_2B_2\right]^2\left[A_2+m_3B_2\right]^2\\
m_3&=&\left[A_3+m_1B_3\right]^2\left[A_2+m_2B_2\right]^2\left[A_2+m_3B_2\right]^2\\
m_4&=&\left[A_1+m_1B_1\right]^6\left[A_1+m_4B_1\right]^2\\
\end{array}}
Here $m_1,m_4$ are the magnetizations of the two different
representative sites in the core and  $m_2,m_3$ are the
magnetizations of the two different representative sites in the
shell. The coefficients in the expanded form of Eq. \re{denk4}
are given by \eq{denk5}{ A^k_p A^l_q B^m_p
B^n_q=\integ{}{}{}dH_iP\paran{H_i}
\cosh^k\paran{J_{p}\nabla}\cosh^l\paran{J_{q}\nabla}\sinh^m\paran{J_{p}\nabla}\sinh^n\paran{J_{q}\nabla}f\paran{H_i,x}|_{x=0}
} where $\nabla$ is the usual differential operator in the
differential operator technique  and the values of indices $p,q$ can
be $p,q=1,2,3$. The function is defined by
\eq{denk6}{f\paran{H_i,x}=\tanh\paran{\beta x + \beta H_i}.} as
usual for the spin-1/2 system. In Eq. \re{denk6}, $\beta=1/(k_B T)$ where $k_B$ is Boltzmann
constant and $T$ is the temperature. The effect of the exponential
differential operator to an arbitrary  function $F(x)$ is given by
\eq{denk7}{\exp{\paran{a\nabla}}F\paran{x}=F\paran{x+a}} with any
constant  $a$. DA will give the results of the Zernike approximation
\cite{refs7} for this system.

With the help of the Binomial expansion, Eq. \re{denk4} can be written in the form
\eq{denk8}{\begin{array}{lcl}
m_1&=&\summ{i=0}{4}{}\summ{j=0}{1}{}\summ{k=0}{2}{}\summ{l=0}{1}{}K_1\paran{i,j,k,l}m_1^im_2^jm_3^k m_4^l\\
m_2&=&\summ{i=0}{1}{}\summ{j=0}{2}{}\summ{k=0}{2}{}K_2\paran{i,j,k}m_1^i m _2^j m_3^k\\
m_3&=&\summ{i=0}{2}{}\summ{j=0}{2}{}\summ{k=0}{2}{}K_3\paran{i,j,k}m_1^i m_2^j m_3^k\\
m_4&=&\summ{i=0}{6}{}\summ{l=0}{2}{}K_4\paran{i,l}m_1^i m_4^l\\
\end{array}} where

\eq{denk9}{\begin{array}{lcl}
K_1\paran{i,j,k,l}&=&\komb{4}{i}\komb{2}{k}A_1^{5-i-l}A_3^{3-j-k}B_1^{i+l}B_3^{j+k}\\
K_2\paran{i,j,k}&=&\komb{2}{j}\komb{2}{k}A_2^{4-j-k}A_3^{1-i}B_2^{j+k}B_3^{i}\\
K_3\paran{i,j,k}&=&\komb{2}{i}\komb{2}{j}\komb{2}{k}A_2^{4-j-k}A_3^{2-i}B_2^{j+k}B_3^{i}\\
K_4\paran{i,l}&=&\komb{6}{i}\komb{2}{l}A_1^{8-i-l}B_1^{i+l}\\
\end{array}}
These coefficients can be calculated from the definitions given in Eq. \re{denk5} with using Eq. \re{denk7}.

For a given Hamiltonian and field distribution parameters, by determining the coefficients  from Eq. \re{denk9} we can obtain a system of coupled non linear equations from Eq. \re{denk8}, and by solving this system we can get the magnetizations $m_i,i=1,2,3,4$. The magnetization of the core $(m_c)$ and shell $(m_s)$ of nanowire, as well as the total magnetization $(m_T)$ can be calculated via
\eq{denk10}{m_c=\frac{1}{7}\paran{6m_1+m_4}, \quad
m_s=\frac{1}{12}\paran{6m_2+6m_3}, \quad
m_T=\frac{1}{19}\paran{6m_1+6m_2+6m_3+m_4}}

Since in the vicinity of the critical point all magnetizations are close to zero, we can obtain another coupled  equation system for determining this critical point by linearizing the equation system given in  Eq. \re{denk8}, i.e.
\eq{denk11}{A.m=0}
where
\eq{denk12}{A=\left(
\begin{array}{cccc}
K_1(1,0,0,0)-1&K_1(0,1,0,0)&K_1(0,0,1,0)&K_1(0,0,0,1)\\
K_2(1,0,0)&K_2(0,1,0)-1&K_2(0,0,1)&0\\
K_3(1,0,0)&K_3(0,1,0)&K_3(0,0,1)-1&0\\
K_4(1,0)&0&0&K_4(0,1)-1\\
\end{array}
\right) }

\eq{denk13}{m=\left(
\begin{array}{c}
m_1\\
m_2\\
m_3\\
m_4\\
\end{array}
\right).}

Critical temperature can be determined from $\mathbf{\mathrm{det(A)=0}}$. As discussed in \cite{ref47}, the matrix $A$ given in Eq. \re{denk12} is invariant under the transformation $J_3\rightarrow -J_3$ then we can conclude that the system with ferromagnetic ($J_3>0$) core-shell interaction has the same critical temperature as that of the system with anti-ferromagnetic ($J_3<0$) core-shell interaction (with the same $|J_3|$) for certain Hamiltonian and magnetic field distribution parameters. Although this discussion has been made for the system with zero magnetic field in \cite{ref47}, this conclusion is also valid for this system, because of the symmetry of the magnetic field distribution. Equation $\mathbf{\mathrm{det(A)=0}}$ is invariant under the transformation $J_3\rightarrow -J_3$ for the nanowire with  magnetic field distribution given in Eq. \re{denk2}.

Beside this, there are some symmetry properties of the coefficients defined by Eq. \re{denk9}. Although the numerical integration in Eq. \re{denk5} do not take too much time, using these symmetry properties will shorten the numerical calculation time.  These symmetry properties can be found in Sec. \ref{app_a}.

\section{Results and Discussion}\label{results}

In this section we discuss the effect of the continous random magnetic field distribution on the phase diagrams of the system. Since the phase diagrams are the same for the $J_3>0$ and $J_3<0$ with the same $|J_3|$, we focus ourselves on the case $J_3>0$, i.e ferromagnetic core-shell interaction. We use the scaled interactions as
\eq{denk14}{J_1=J, \quad r_n=\frac{J_n}{J}, \quad n=2,3.}
We start with a single Gaussian magnetic field distribution.

\subsection{Single Gaussian Distribution}

The form of single Gaussian distribution which is defined in Eq.
\re{denk2} is governed by only one parameter $\sigma$ , which is the
width of the distribution. This distribution distributes negative
and positive valued magnetic fields -which are chosen from the
Gaussian distribution- to lattice sites so that sum of all lattice
site's magnetic field is equal to zero. Although the total magnetic
field is zero, randomly distributed negative and positive fields
drag the system to the disordered phase. On the other hand, the
interactions $J_n,(n=1,2,3)$  enforce the system to stay in the
ordered phase. Another factor is the temperature which causes
thermal agitations which induce disordered phase when energy
supplied by the temperature to the system is high enough. Thus a
competition takes place between these factors.

\begin{figure}[h]\begin{center}
\epsfig{file=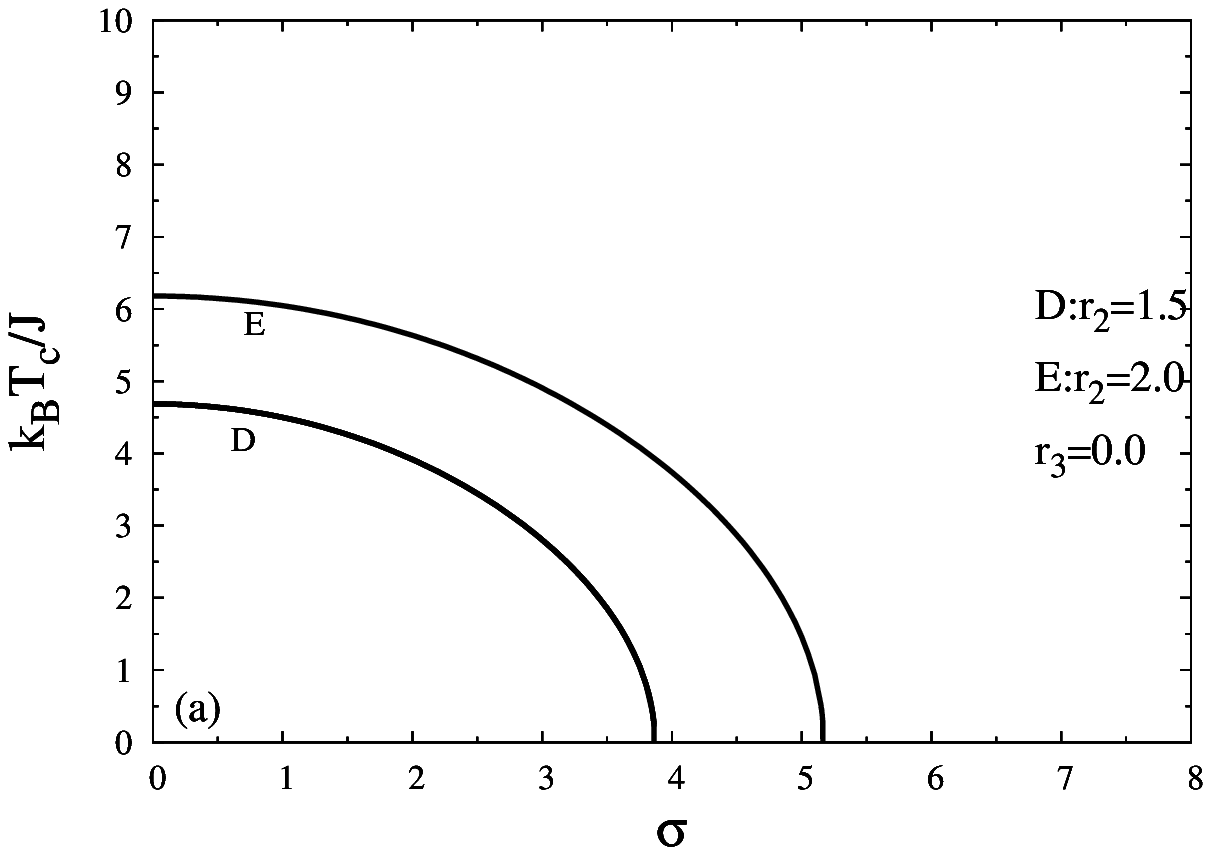, width=4.5cm}
\epsfig{file=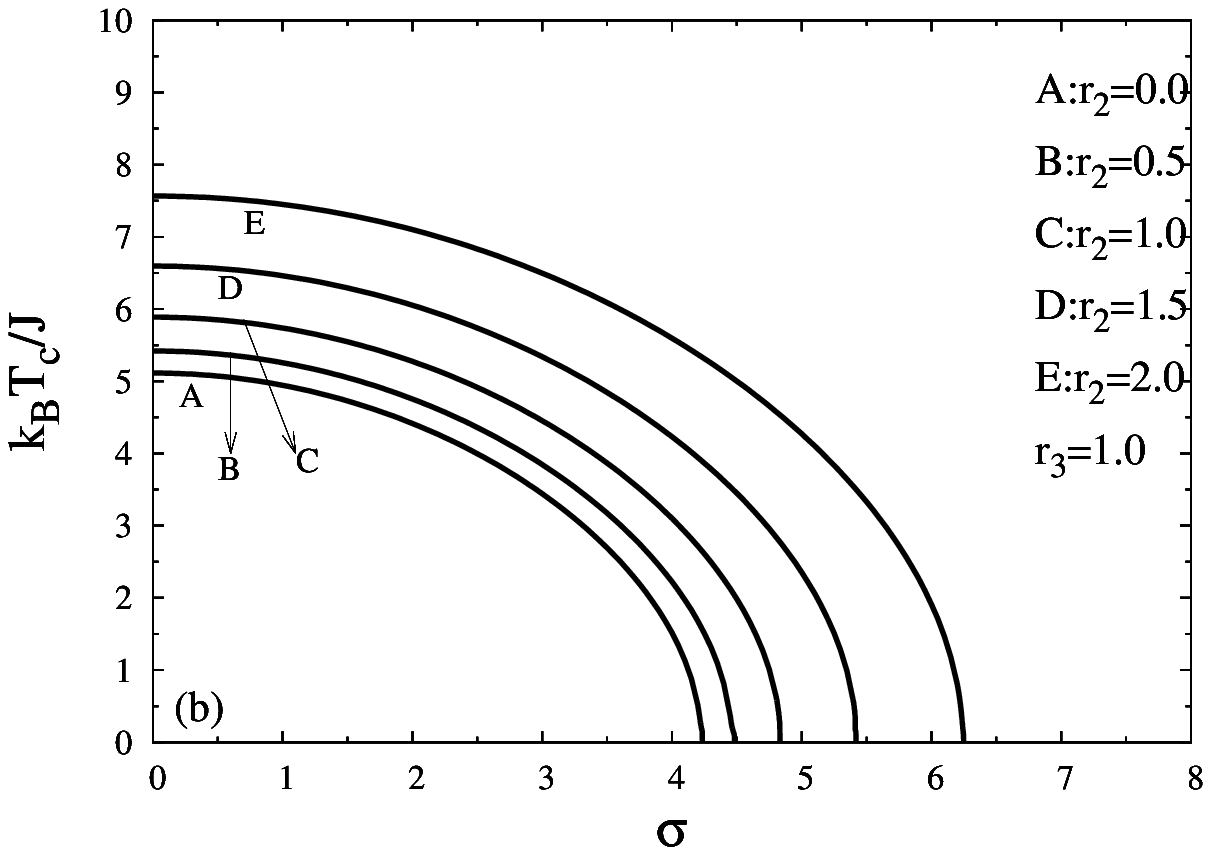, width=4.5cm}
\epsfig{file=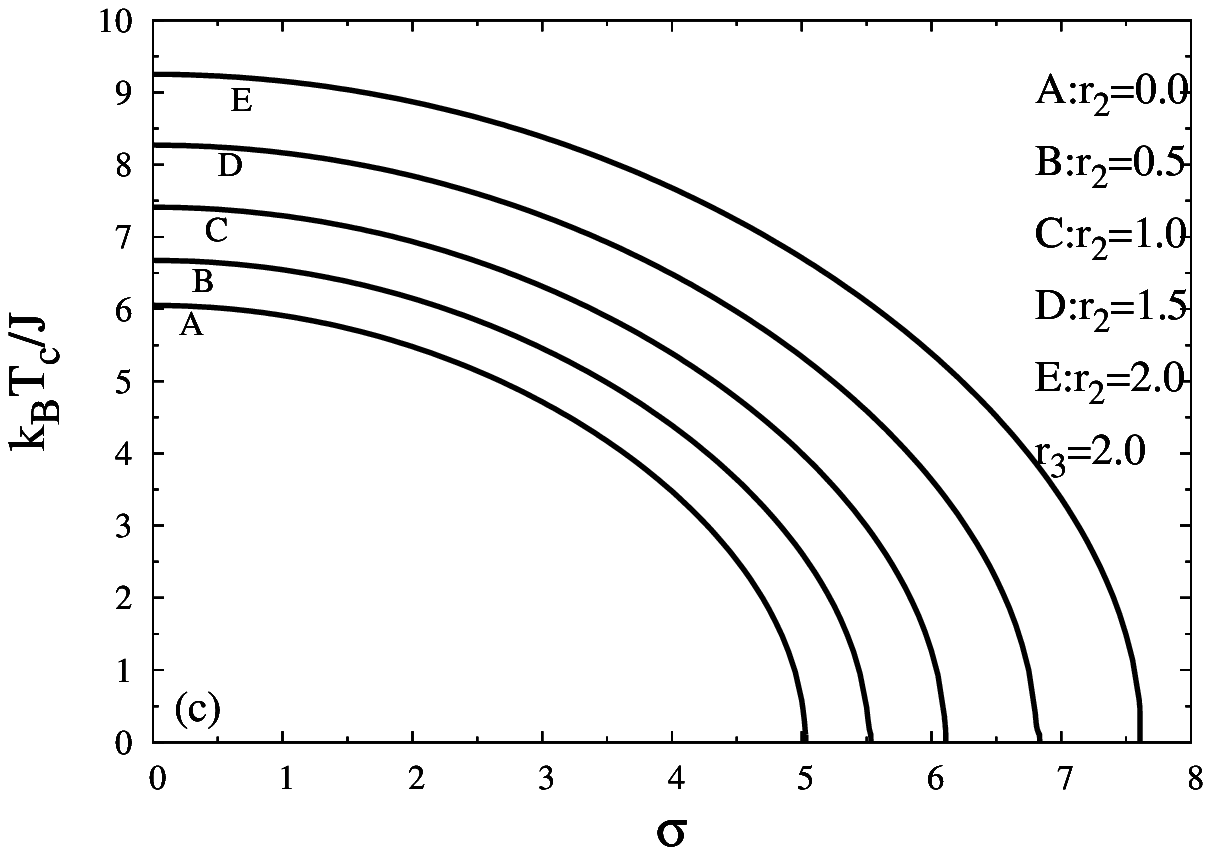, width=4.5cm}
\end{center}
\caption{The phase diagrams of the nanowire  with single Gaussian random
field distribution in the ($k_BT_c/J, \sigma)$ plane for different
$r_2,r_3$ values.} \label{sek2}\end{figure}

The phase diagrams in $(k_BT_c/J, \sigma)$ plane can be seen in Fig.
\ref{sek2} for different $r_2$ and $r_3$ values. Firstly we can see
from Fig. \ref{sek2} that the general effect of increasing $\sigma$
is to decrease the critical temperature, as expected. There is not any evidence of
reentrant behavior in the phase diagrams, as in ordinary Ising
lattices\cite{refs9}. The critical temperature value at $\sigma=0$
and the critical value of the $\sigma$ -which makes the critical
temperature zero- depend on $r_2,r_3$, i.e. when $r_2$ or $r_3$
raises, these two critical values gets higher except one region. As
in the same system with discrete distributions \cite{refs8} when
$r_3=0.0$ up to a certain value  of the $r_2$, the phase diagrams
are the same in $(k_BT_c/J, \sigma)$ plane. In Fig. \ref{sek2}(a)
all phase diagrams are the same for $r_2 \le 1.5$. This fact can be
explained by taking into account the total spin-spin interaction
strengths of the core and shell which do not interact with each other
when $r_3=0.0$.

\begin{figure}[h]\begin{center}
\epsfig{file=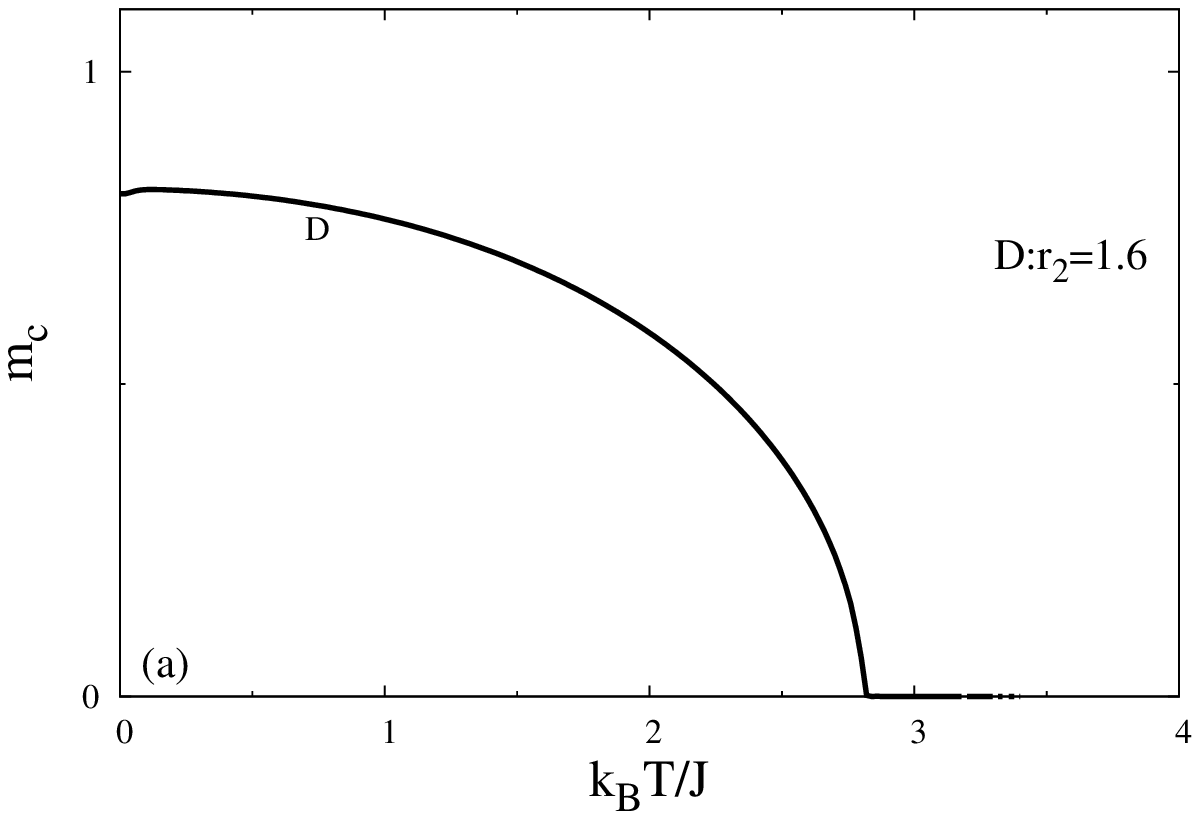, width=4cm,height=4cm}
\epsfig{file=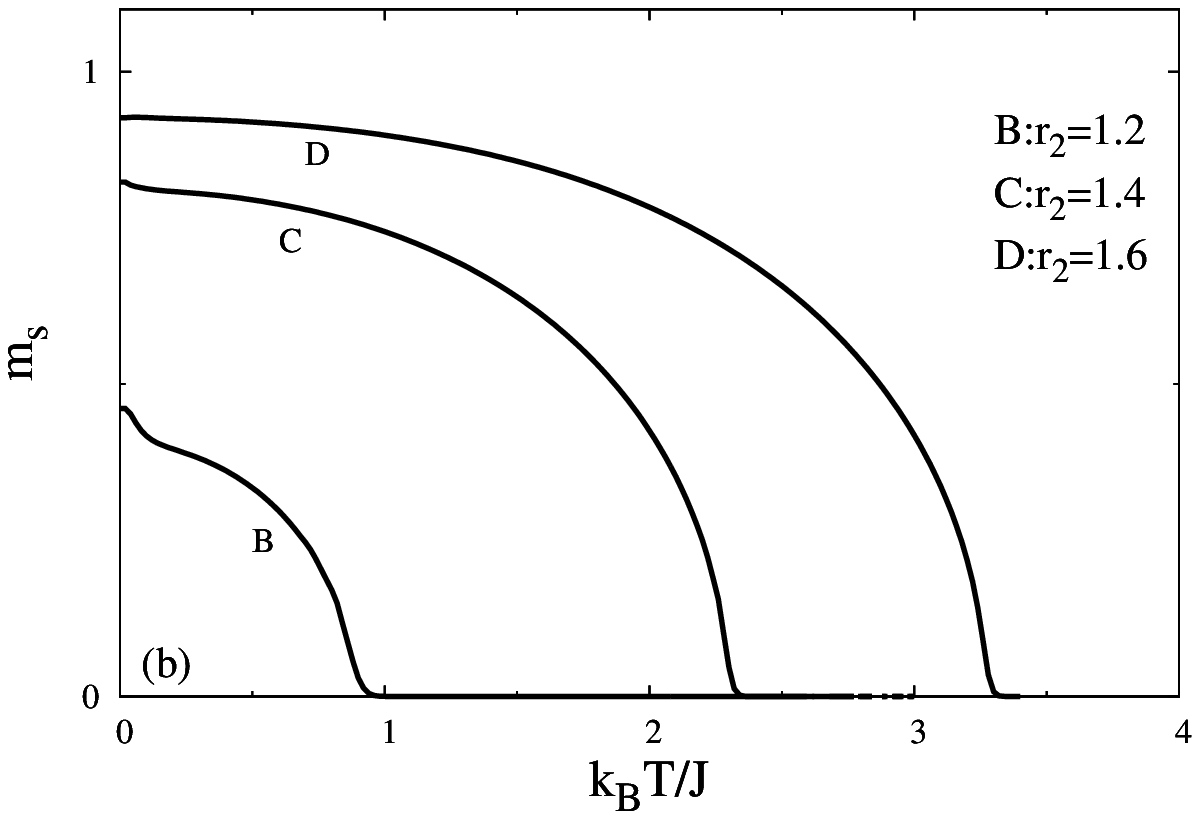, width=4cm,height=4cm}
\epsfig{file=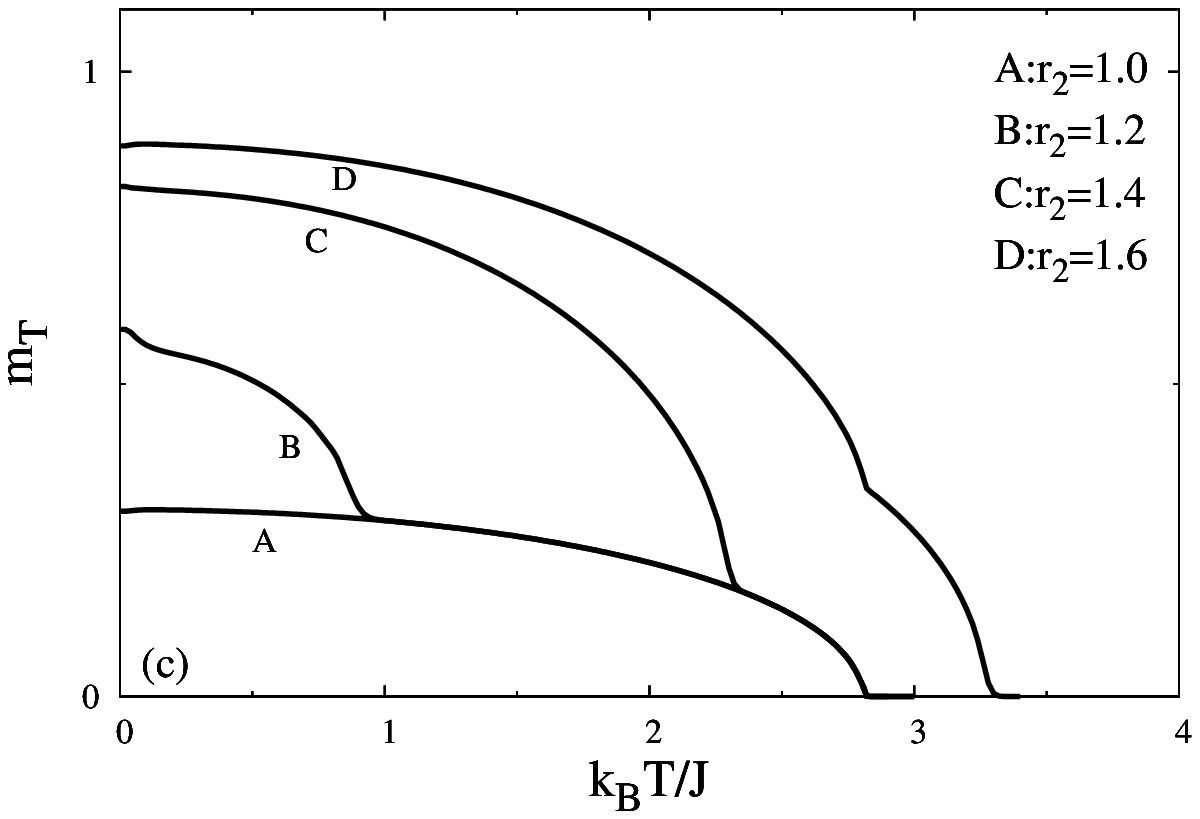, width=4cm,height=4cm}
\end{center}
\caption{Variation of the magnetization with temperature for some
selected $r_2$ values of the nanowire with single Gaussian random
field distribution. The fixed parameter values  are
$r_3=0.0,\sigma=3.0$.} \label{sek3}\end{figure}

In the absence of the interaction between the core and shell, the
critical temperature of the system is determined by the core, up to a
certain $r_2>1.0$ value. Two distinct shell spins (which have
magnetizations $m_2,m_3$) have four nearest neighbors, and two
distinct core spins (magnetizations of which are labeled as
$m_1,m_4$) have five and eight nearest neighbors, respectively. Thus
at the value of $r_2=1.0$, core has higher critical temperature than
the shell. As $r_2$ increases, the critical temperature of the shell
gets higher and after a certain value of $r_2$\textbf{,} the shell begins
to determine the critical temperature of the system.  Before this
value, changing $r_2$ can not change the phase diagrams of the
system. The dependence of the critical temperature values of the
core and shell can be seen in variation of their magnetization curves with
temperature which are given in  Fig. \ref{sek3} for some selected
values of $r_2$ with fixed $r_3=0.0$ and $\sigma=3.0$ values.

Within this region, although increasing $r_2$ does not change the
critical temperature of the system for a fixed value of $\sigma$, it
causes change in the magnetization behavior with the temperature as
seen in Fig. \ref{sek3}(c). As seen in Fig. \ref{sek3}(b), as
$r_2$ increases for a fixed $\sigma$, the ground state magnetization
and the critical temperature of the shell layer increase. When $r_2$ reaches the
value that makes the critical temperature of the shell equal to the
critical temperature of the core, increasing $r_2$ after this value
manifests itself in rising critical temperature of the system.

We can see from Fig. \ref{sek2} that, increasing $r_2$ and $r_3$
values make the ferromagnetic region wider in ($k_BT_c/J, \sigma$)
plane (except for the region explained above), since increasing $r_2$ and $r_3$ values raises the absolute
value of the lattice energy coming from the spin-spin interaction,
which must be overcomed by both thermal agitations and random
magnetic fields for the phase transition from an ordered phase to a
disordered one.


\begin{figure}[h]\begin{center}
\epsfig{file=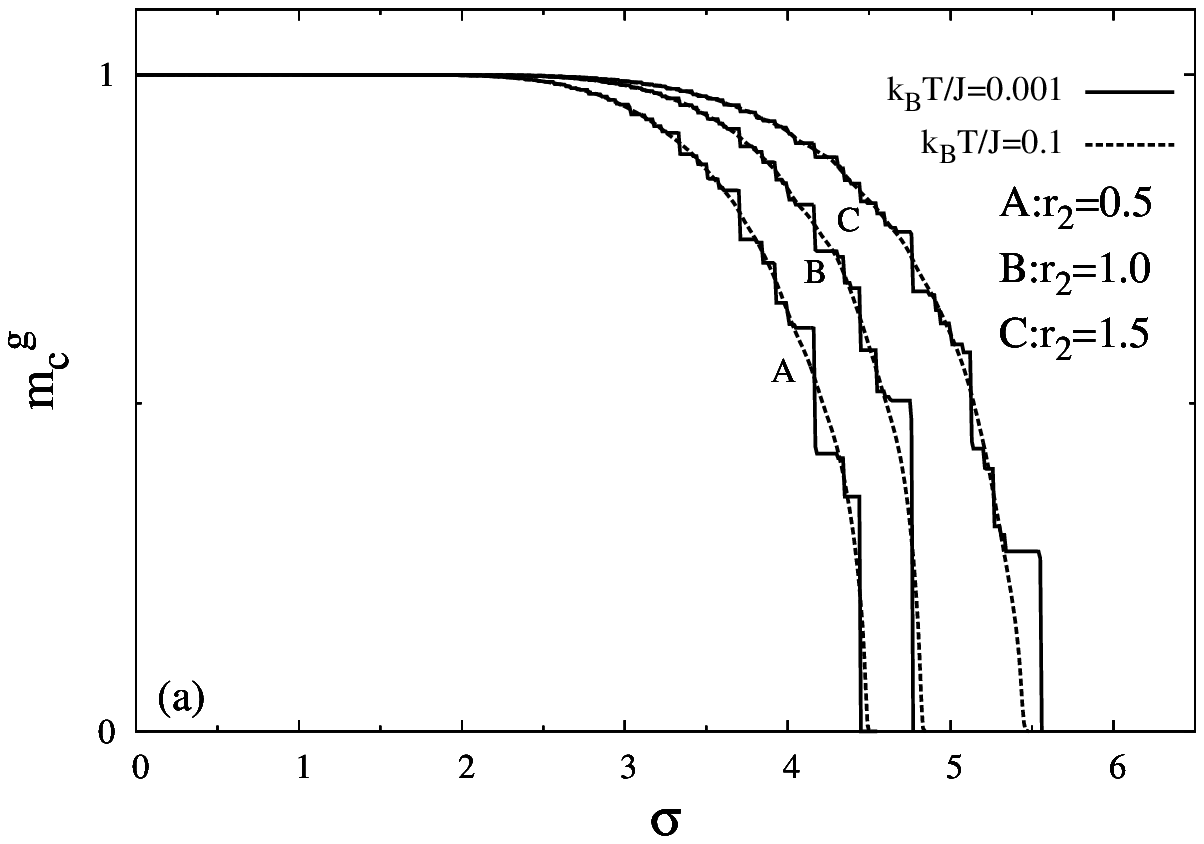, width=4.5cm,height=4cm}
\epsfig{file=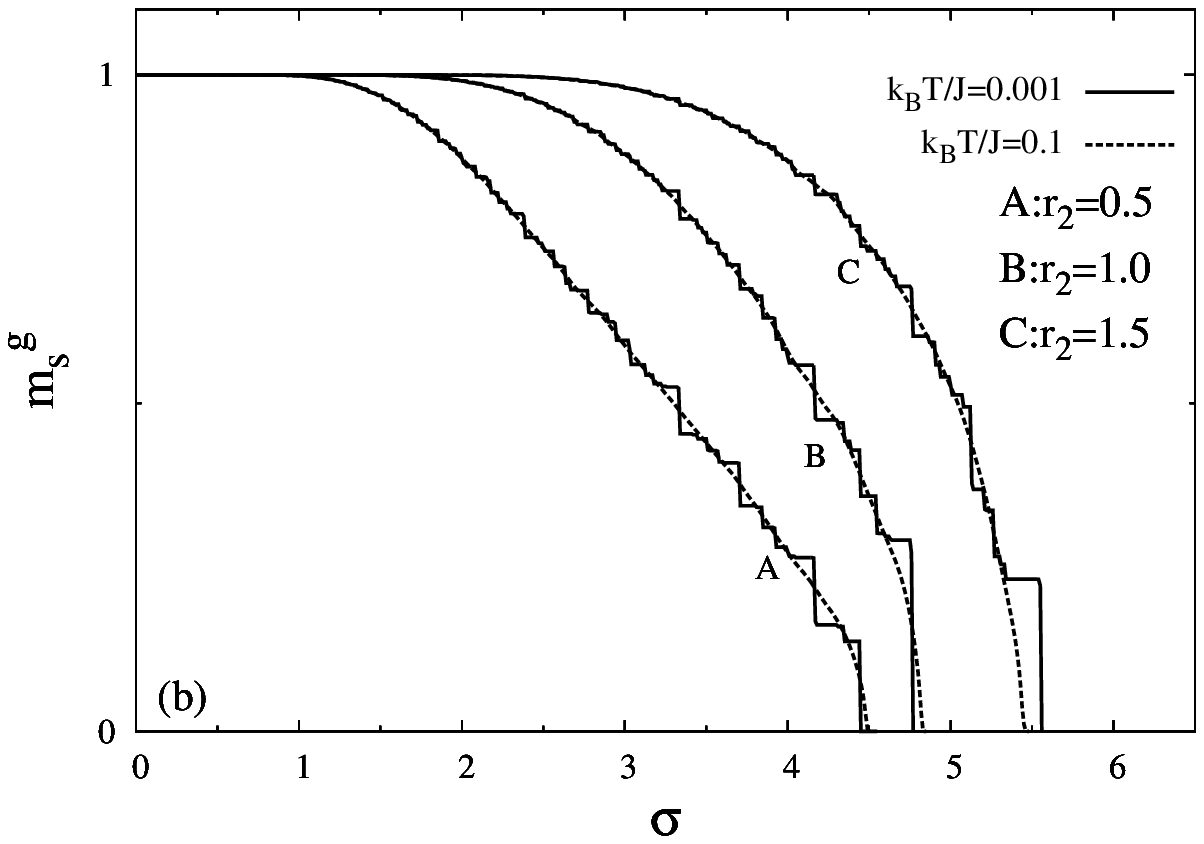, width=4.5cm,height=4cm}
\epsfig{file=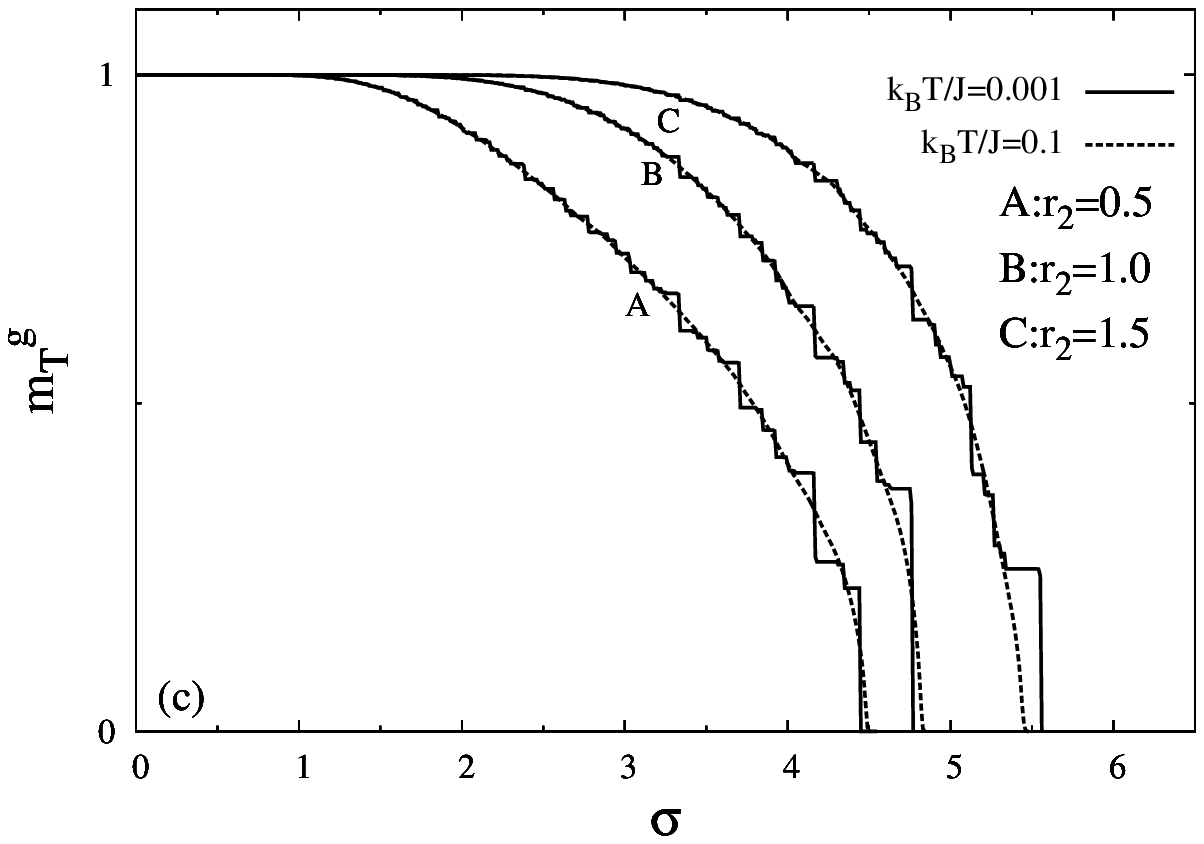, width=4.5cm,height=4cm}
\end{center}
\caption{Variation of the ground state magnetizations with $\sigma$
for some selected values of $r_2$ for the system with single
Gaussian magnetic field distribution. Each magnetization calculated
at a temperature $k_BT/J=0.001$ can be regarded as ground
state of the system. At the same time, the magnetization values
calculated at the temperature $k_BT/J=0.1$ are plotted as dashed lines
with the same $r_2$ values.  Fixed parameter value is $r_3=1.0$.}
\label{sek4}\end{figure}

As we done in our earlier work \cite{refs8} for discrete
distributions, now we want to investigate the variation of ground state
magnetization behaviors with $\sigma$. Then our question is; how the ground state magnetizations change as $\sigma$ is varied for different $r_2,r_3$ values? We know from the conclusions of our
earlier work \cite{refs8}, bimodal and trimodal discrete random
magnetic field distributions produces a number of ground
state magnetizations which are different from zero (disordered
phase) and one (ordered phase). These intermediate states may be
regarded as partially ordered phases. In Fig. \ref{sek4}, we plot the
variation of the ground state magnetization values ($m_c^g,m_s^g,m_T^g$
core, shell and total ground state magnetization, respectively) as a function of
the width of the single Gaussian distribution ($\sigma$) for some
selected values of $r_2$ with $r_3=1.0$. We can see from the Fig.
\ref{sek4} that, this distribution creates higher number of these
partially ordered phases, in comparison with the discrete distributions. Although
increasing $\sigma$ decreases the critical temperature of the system
gradually as seen in Fig. \re{sek2}, EFT based on DA approximation gives
that, increasing $\sigma$ decreases the ground state magnetization
as a general trend, but it creates some plateaus i.e. there are some
regions where the ground state magnetizations does not change with varying $\sigma$ values.
With increasing temperature, the effect of the
increasing $\sigma$ on the magnetization becomes continuous i.e.
increasing $\sigma$ decreases the magnetization continuously at higher temperatures as seen again in Fig. \ref{sek4}.

\subsection{Double Gaussian Distribution}

\begin{figure}[h]\begin{center}
\epsfig{file=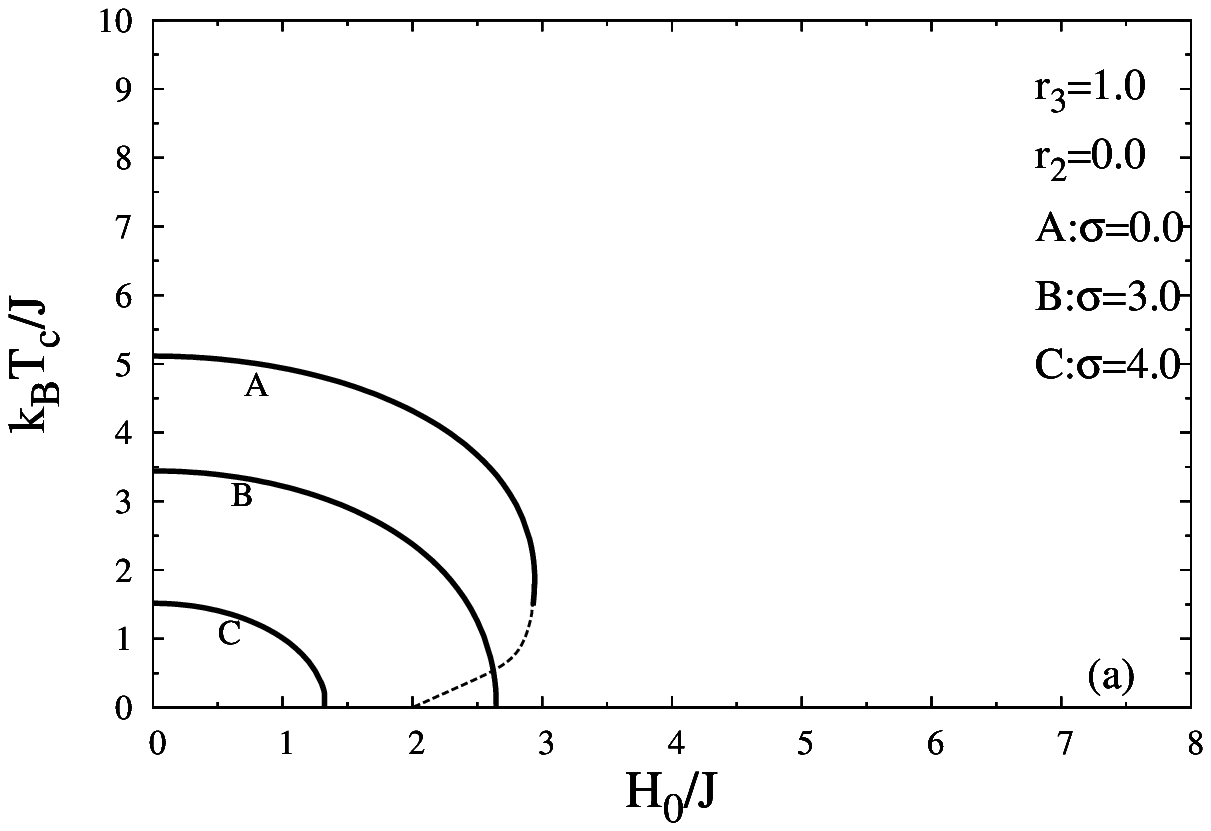, width=6cm}
\epsfig{file=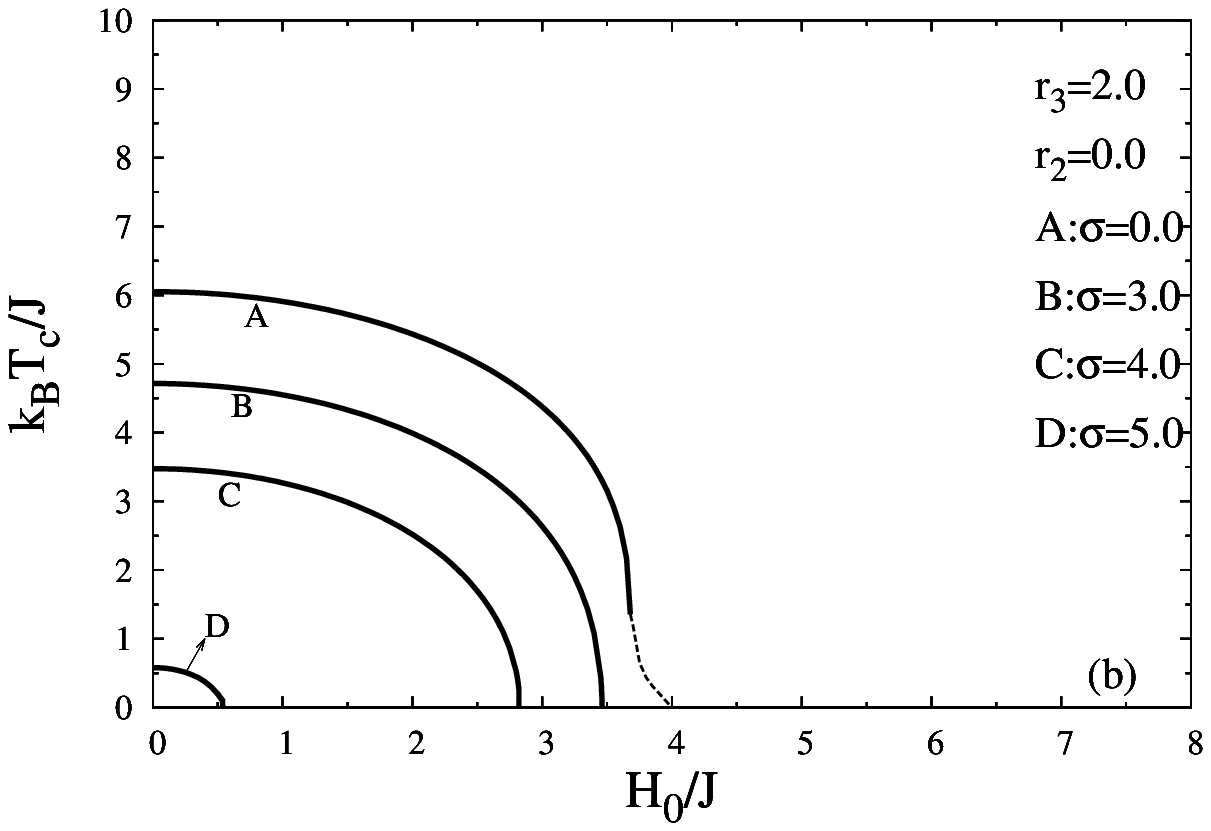, width=6cm}

\epsfig{file=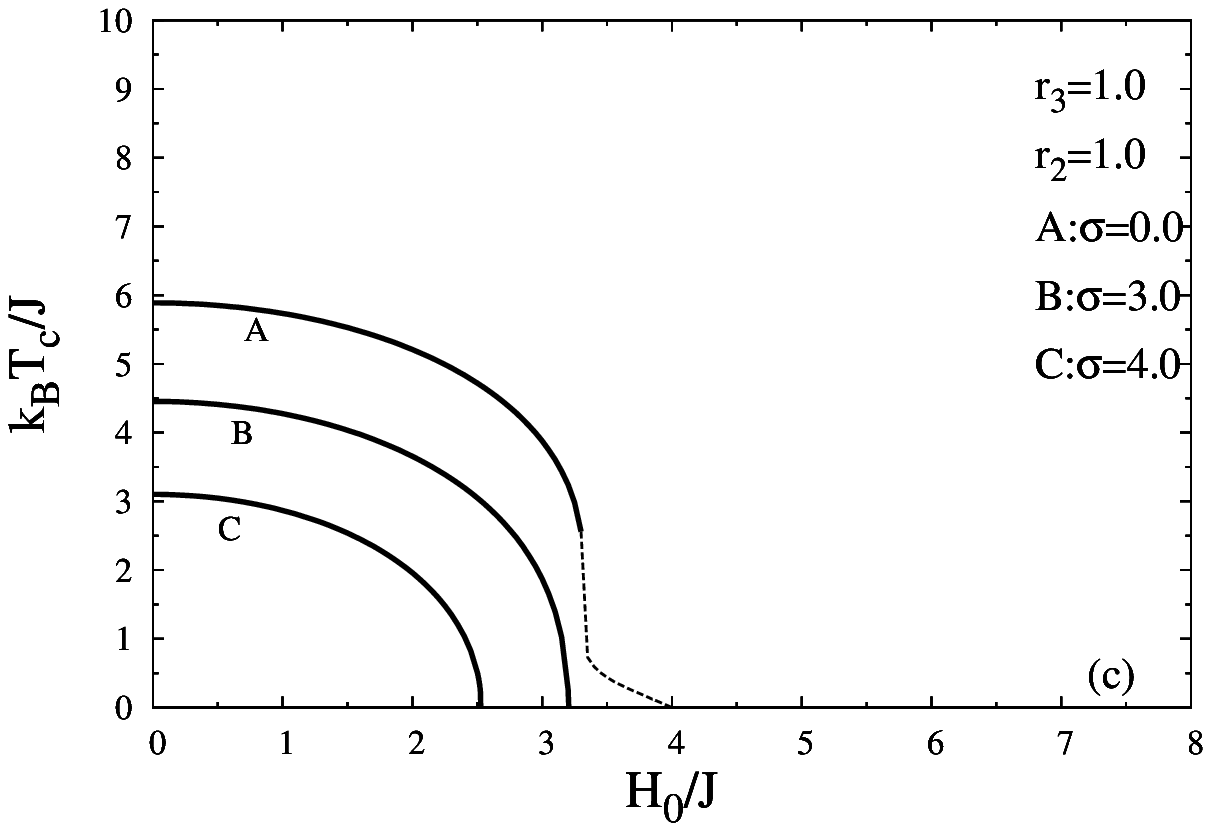, width=6cm}
\epsfig{file=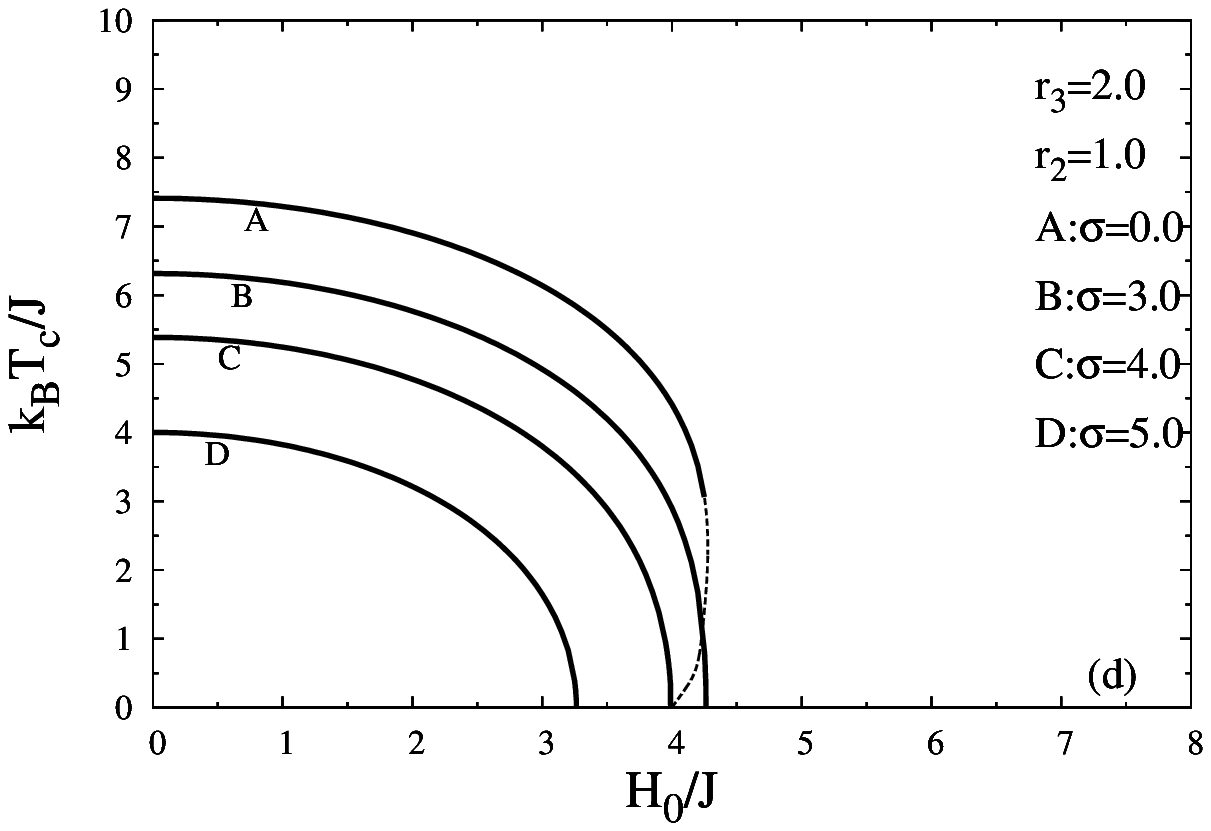, width=6cm}

\epsfig{file=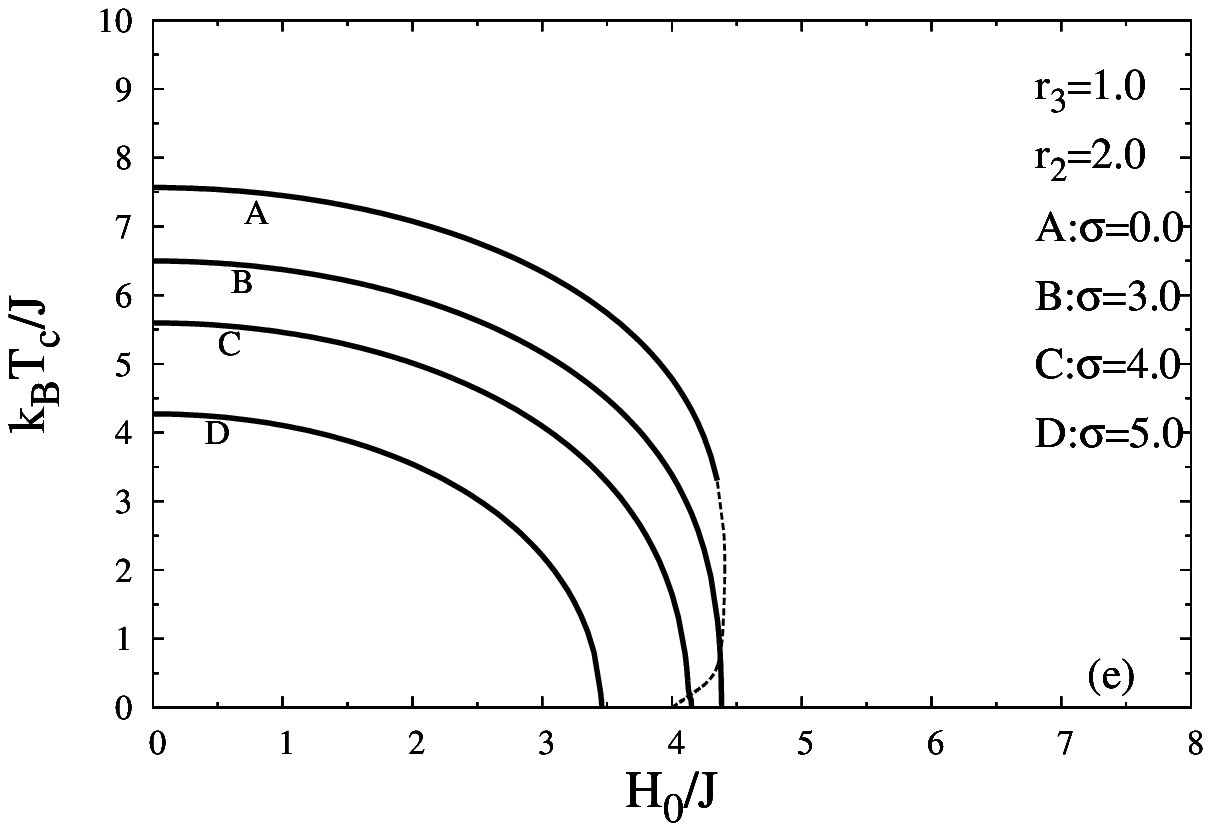, width=6cm}
\epsfig{file=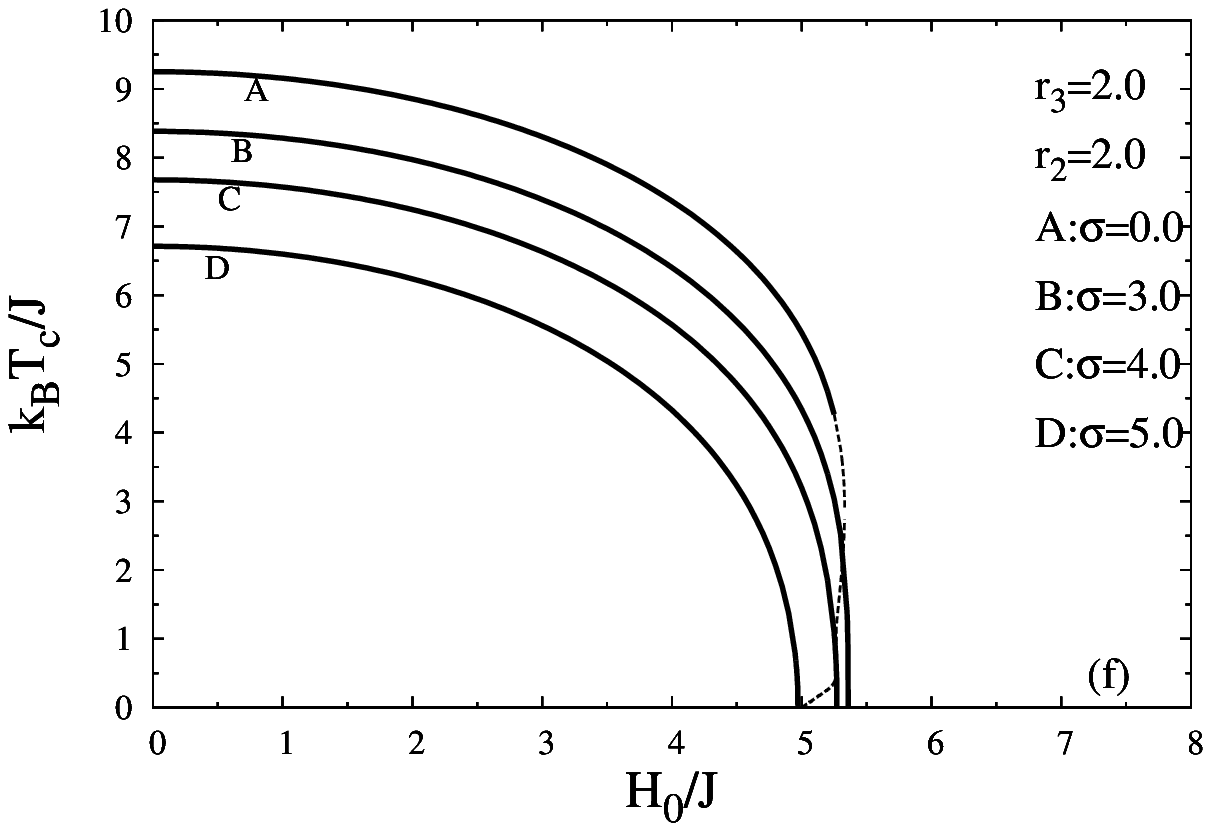, width=6cm}

\end{center}
\caption{The phase diagrams of the nanowire  with double Gaussian random field distribution in the $(k_BT_c/J, H_0/J)$ plane for different $r_2,r_3,\sigma$ values}
\label{sek5}\end{figure}

Let us investigate the phase diagrams of the nanowire  with double Gaussian distribution. This distribution is given by Eq. \re{denk2} with $p=0$.
In Fig. \ref{sek5}, we plot the phase diagrams in the $(k_BT_c/J, H_0/J)$ plane for some selected $r_2,r_3$ and $\sigma$ values. First, we see from the Fig. \ref{sek5} that, for fixed $r_2,r_3$ and $\sigma$ values, when the distance between the centers of the Gaussians ($2H_0/J$) increases, the critical temperature of the system gets smaller. Beside this, for fixed $r_2,r_3$ values, increasing $\sigma$  makes the ferromagnetic region in the $(k_BT_c/J, H_0/J)$ plane narrower, since the randomness increases due to increasing $\sigma$. Another effect of the rising $\sigma$ in this distribution is that, increasing $\sigma$ destroys the reentrant behavior which exist in the system for given $r_2,r_3$ values for the bimodal distribution (i.e. double Gaussian distribution with $\sigma=0.0$) as seen in Figs. \ref{sek5}(a),(d),(e),(f). Not only reentrant behavior disappears when $\sigma$ rises, but also first order transitions and tricritical points. At the same time as seen in Figs. \ref{sek5}(b) and (c), rising $\sigma$ also destroys other first order transitions which originate from an ordered phase to a disordered phase and which may not come with the reentrant behavior. Then  we can say that for the nanowire, while bimodal distribution can induce a first order transitions -and also tricrtical points-  at higher $H_0/J$ values, double Gaussian distrbitubion can destroy these first order transitions when $\sigma$ is large enough.

\begin{figure}[h]\begin{center}
\epsfig{file=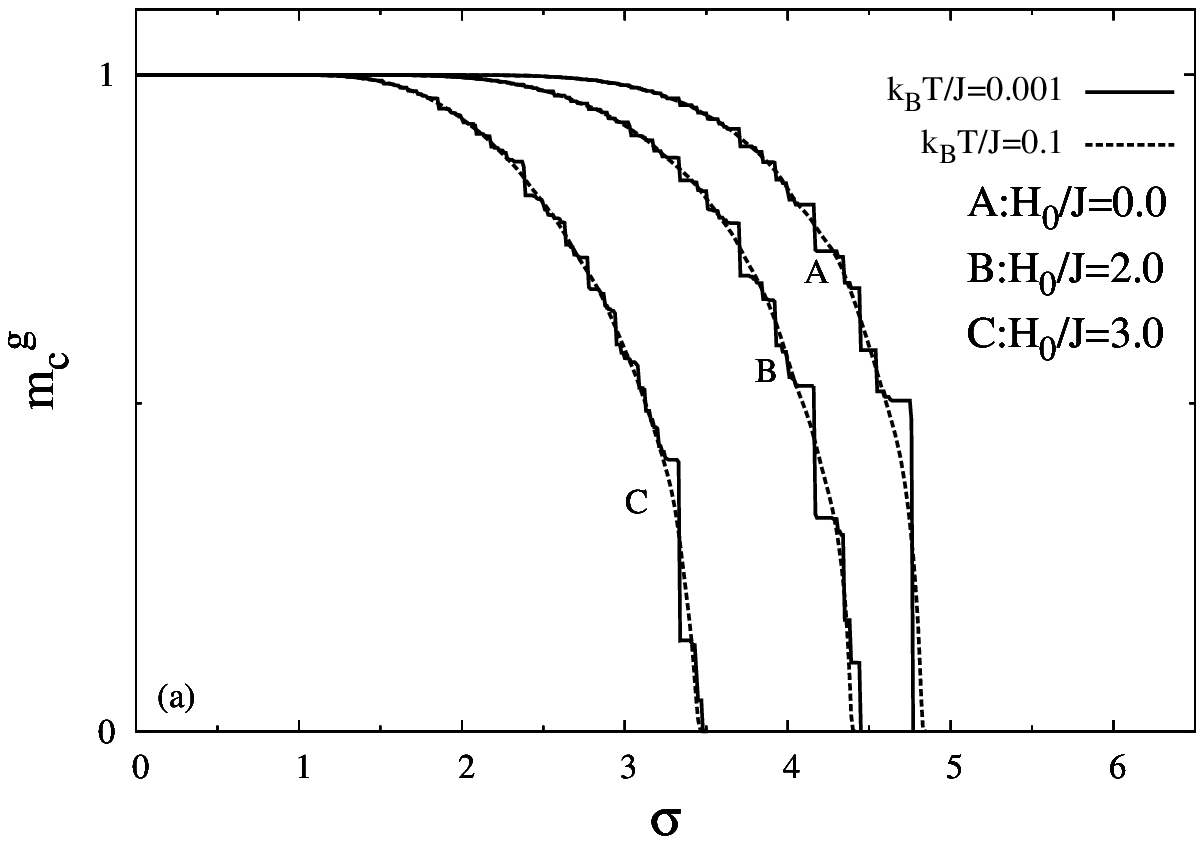, width=4.5cm,height=4cm}
\epsfig{file=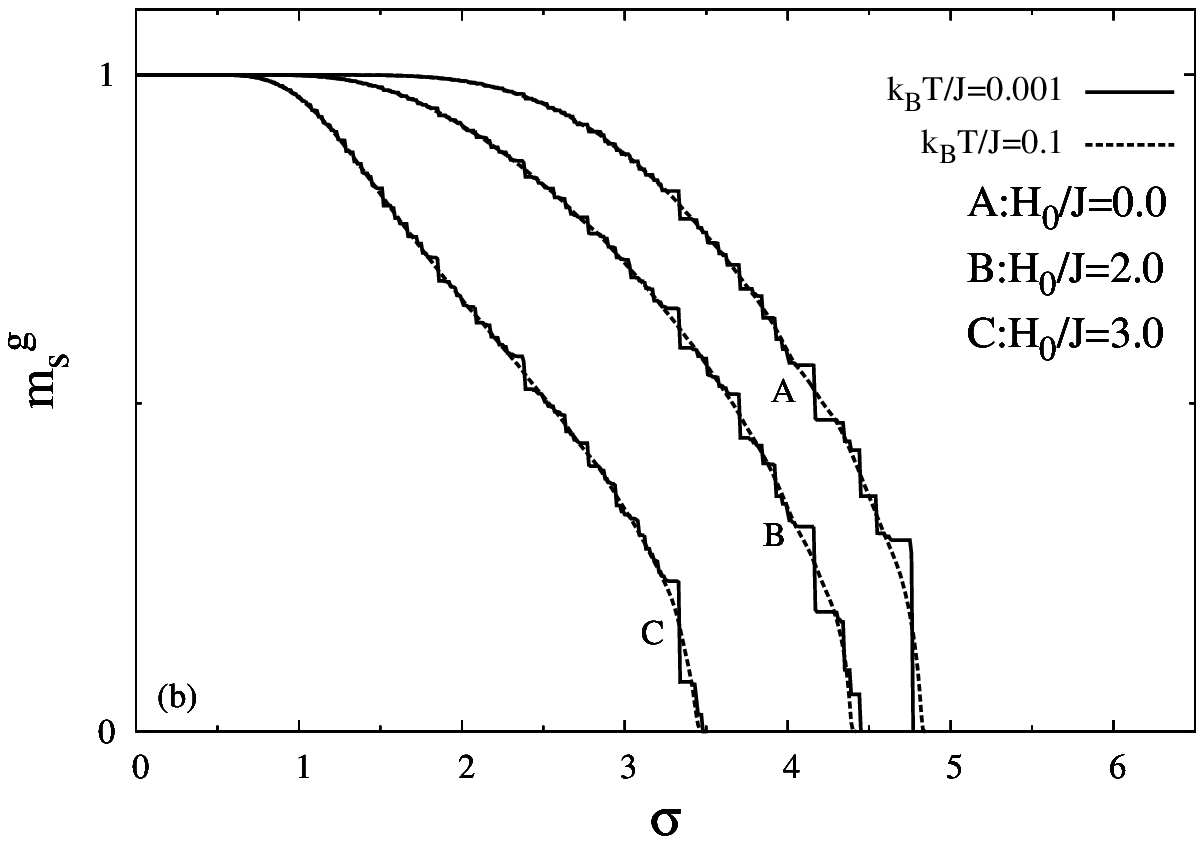, width=4.5cm,height=4cm}
\epsfig{file=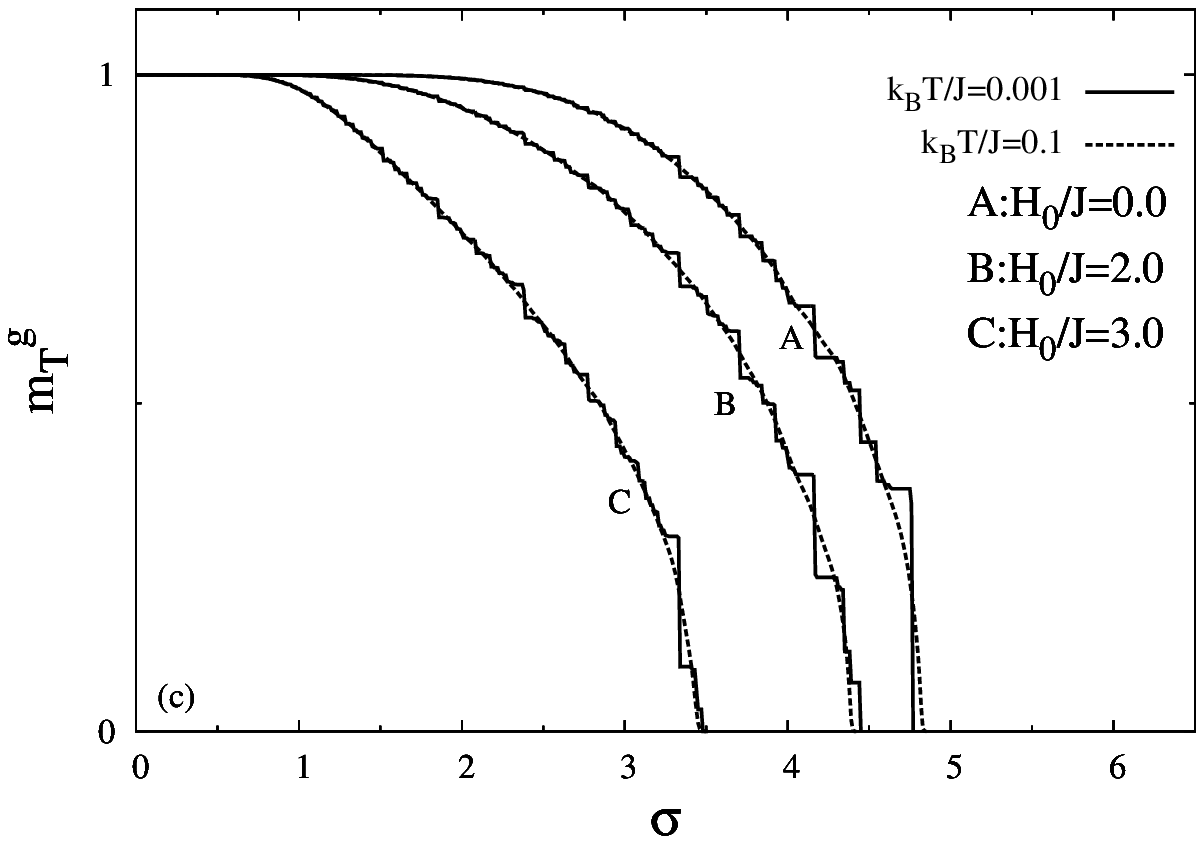, width=4.5cm,height=4cm}
\end{center}
\caption{
Variation of the ground state magnetizations with $\sigma$
for some selected values of $H_0/J$ for the system with double
Gaussian magnetic field distribution. Each magnetization calculated
at a temperature $k_BT/J=0.001$ can be regarded as ground
state of the system. At the same time, the magnetization values
calculated at the temperature $k_BT/J=0.1$ are plotted as dashed lines
with the same $H_0/J$ values.  Fixed parameter values are $r_2=1.0$ and $r_3=1.0$.
} \label{sek6}\end{figure}

The ground state magnetizations appear in  more complicated forms for
this distribution than those corresponding to the bimodal distribution\cite{refs8}. This can be seen
from Fig. \ref{sek6} which represents the ground state magnetizations corresponding to the phase diagrams given in Fig. \ref{sek5}(c).
As in the single Gaussian distribution, large number of partially ordered phases appear
in the system which disappear when the temperature rises.

\section{Conclusion}\label{conclusion}

In this work, the effect of the continuous  random magnetic field distributions on the phase
diagrams and ground state magnetizations of the Ising nanowire has been investigated.

There are two distribution parameters ($H_0/J$ and $\sigma$) of the random magnetic fields which can also be regarded as degree of the randomness which is imposed by the distribution of the magnetic field on the nanowire. Rising  $H_0/J$ or $\sigma$ can be considered as higher degree of randomness. This shows itself as a change in the phase diagrams which are plotted in $(k_BT_c/J-\sigma)$ or $(k_BT_c/J-H_0/J)$ planes. We can say that, this change is two fold when we compare the results of the continuous distribution with the discrete distribution. First one is obvious\textbf{;}  the ferromagnetic region in that planes
collapses with rising degree of randomness. The second one is, rising degree of randomness imposed by rising $\sigma$ on the system, destroys all first order transition lines and also tricritical points at which the first order and second order transition lines meet. Rising degree of randomness which is caused by rising $\sigma$, also destroys reentrant behavior. While the system with magnetic field distribution given in Eq. \re{denk2} with $\sigma=0.0$ can show reentrant behavior, after a certain $\sigma$ -which depends on $r_2$ and $r_3$- this behavior disappears. This means that, the system with double Gaussian distribution can not pass the border of the two phases with thermal agitations from a disordered phase to an ordered phase after a certain $\sigma$ value.

Rising degree of randomness also shows itself in the ground state magnetizations. As one can see in \cite{refs8}, bimodal magnetic field distribution can induce a few plateaus in the $(m_T^g-K_BT/J)$ plane. These plateaus where the ground state magnetization values do not change with rising randomness can be regarded as partially ordered phases. We can see from Figs. \ref{sek4} and \ref{sek6} that, continuous random magnetic field distribution can induce higher number of these partially ordered phases.   Also it can be seen in Fig. \ref{sek4} and \ref{sek6} that, it is impossible  to relate the magnetization values or widths of the constant ground state magnetization region of the partially ordered phases, with the system parameters.

When the temperature  varies a little, these partially ordered phases disappear. This means that there have to be some first order transition lines ending with isolated critical points in Figs. \ref{sek2} and  \ref{sek5}. However in order to clarify whether this partially ordered phases are artifact or not of EFT based on DA, it is necessary to investigate the problem with more advanced methods.

We hope that the results  obtained in this work may be beneficial form both theoretical and experimental point of view.

\section{Appendix A: Symmetry properties of the coefficients}\label{app_a}

As seen in the definitions of the the coefficients given by
\re{denk9}, interchanging between certain indices do not change the
part of that coefficient which is defined by \re{denk5}. For
obtaining the simple symmetry properties originated from this fact,
let us start with the first coefficient by defining\eq{denk_app1}{
K_1^\prime\paran{i,j,k,l}=\left[\komb{4}{i}\komb{2}{k}\right]^{-1}K_1\paran{i,j,k,l}=A_1^{5-i-l}A_3^{3-j-k}B_1^{i+l}B_3^{j+k}.
} From the definition given in Eq. \re{denk5} we can see that Eq.
\re{denk_app1} is invariant  under the transformations $i\rightarrow
l, l\rightarrow i $ and $j\rightarrow k, k\rightarrow j$, thus we
can use this property as \eq{denk_app2}{\begin{array}{lcl}
K_1\paran{i,j,k,l}&=&\komb{4}{i}\komb{2}{k}K_1^\prime\paran{i,j,k,l}\\
K_1\paran{l,j,k,i}&=&\komb{4}{l}\komb{2}{k}K_1^\prime\paran{i,j,k,l}\\
K_1\paran{i,k,j,l}&=&\komb{4}{i}\komb{2}{j}K_1^\prime\paran{i,j,k,l}\\
K_1\paran{l,k,j,i}&=&\komb{4}{l}\komb{2}{j}K_1^\prime\paran{i,j,k,l}\\
\end{array}}
i.e.,
 \eq{denk_app3}{\begin{array}{lcl}
K_1\paran{l,j,k,i}&=&\komb{4}{l}\komb{4}{i}^{-1}K_1\paran{i,j,k,l}\\
K_1\paran{i,k,j,l}&=&\komb{2}{j}\komb{2}{k}^{-1}K_1\paran{i,j,k,l}\\
K_1\paran{l,k,j,i}&=&\komb{4}{l}\komb{2}{j}\komb{4}{i}^{-1}\komb{2}{k}^{-1}K_1\paran{i,j,k,l}\\
\end{array}}
Eq. \re{denk_app3} reduces the total number of coefficients which
has to be calculated as separately from 60 to 45. In a similar way, for
the other coefficients we can obtain similar symmetry properties as
\eq{denk_app4}{ K_2\paran{i,k,j}=K_2\paran{i,j,k} }

\eq{denk_app5}{ K_3\paran{i,k,j}=K_3\paran{i,j,k} }

\eq{denk_app6}{ K_4\paran{l,i}=\komb{6}{l}\komb{2}{i}\komb{6}{i}^{-1}\komb{2}{l}^{-1}K_4\paran{i,l} }

These properties reduce the number of separate coefficients from 18 to 6,
27 to 9, 21 to 15, respectively. Thus for this system, total number
of coefficients reduces from 126 to  75.

On the other hand, another important symmetry property of the
coefficients which is useful in the calculations of the integration
given in Eq. \re{denk5} is about the magnetic field. Let us choose
$P\paran{H_i}=\delta\paran{H_i-H_0}$ as a magnetic field
distribution and denote the coefficient defined by Eq. \re{denk5} as
$\Theta_{klmn}\paran{H_0,J_p,J_q}=A^k_p A^l_q B^m_p B^n_q$. By
writing hypergeometric functions in terms of the exponentials then
writing them with binomial distribution and using Eq. \re{denk7}
we get \eq{denk_app7}{
\Theta_{klmn}\paran{H_0,J_p,J_q}=\paran{\frac{1}{2}}^{k+l+m+n}\summ{r=0}{k}{}\summ{s=0}{l}{}\summ{t=0}{m}{}\summ{v=0}{n}{}
C_{rstv}f\paran{H_0,a_pJ_p+a_qJ_q} } where \eq{denk_app8}{
C_{rstv}=\komb{k}{r}\komb{l}{s}\komb{m}{t}\komb{n}{v}\paran{-1}^{m+n-t-v}
} and \eq{denk_app9}{ a_p=2r+2t-m-k, \quad a_q=2s+2v-l-n } Now we
can see from Eq. \re{denk_app7} that, for each term, $C_{rstv}f\paran{H_0,a_pJ_p+a_qJ_q}$ has a corresponding term which is
$C_{k-r,l-s,m-t,n-v}f\paran{H_0,-a_pJ_p-a_qJ_q}$ in the sum. It can
be seen from Eq. \re{denk_app8} that, these two coefficient are related to each other
by \eq{denk_app10}{ C_{k-r,l-s,m-t,n-v}=\paran{-1}^{m+n}C_{rstv} }
which means that for odd $m+n$, the terms which have
$f\paran{H_0,a_pJ_p+a_qJ_q}$ and $f\paran{H_0,-a_pJ_p-a_qJ_q}$ are
opposite signed coefficients. If  $m+n$ is even then the terms which have
$f\paran{H_0,a_pJ_p+a_qJ_q}$ and $f\paran{H_0,-a_pJ_p-a_qJ_q}$ will
be the same signed coefficients in the sum given in Eq. \re{denk_app7}. Thus we can conclude that,

\eq{denk_app11}{ \Theta_{klmn}\paran{H_0,-J_p,-J_q}=\left\{
\begin{array}{lcl}
\Theta_{klmn}\paran{H_0,J_p,J_q}, &\quad& \text{m+n even}\\
-\Theta_{klmn}\paran{H_0,J_p,J_q}, &\quad& \text{m+n odd}\\
\end{array}
\right. }

Since the functions defined in Eq. \re{denk6} have the property
$f\paran{H_i,x}=-f\paran{-H_i,-x}$, then we can write Eq.
\re{denk_app11} as

\eq{denk_app12}{  \Theta_{klmn}\paran{-H_0,J_p,J_q}=\left\{
\begin{array}{lcl}
-\Theta_{klmn}\paran{H_0,J_p,J_q}, &\quad& \text{m+n even}\\
\Theta_{klmn}\paran{H_0,J_p,J_q}, &\quad& \text{m+n odd}\\
\end{array}
\right. }

Since the distribution given in Eq. \re{denk2} is symmetric about
$H_0=0$, then in integration in Eq. \re{denk5} for any term
$\Theta_{klmn}\paran{H_0,J_p,J_q}$ has a corresponding
$\Theta_{klmn}\paran{-H_0,J_p,J_q}$ term.  This means that in the
case where random field distribution is given by Eq. \re{denk2},

\eq{denk_app13}{ \Theta_{klmn}\paran{J_p,J_q}=\integ{}{}{}dH_0
P\paran{H_0}\Theta_{klmn}\paran{H_0,J_p,J_q}\left\{
\begin{array}{lcl}
=0, &\quad& \text{m+n even}\\
\ne 0, &\quad& \text{m+n odd}\\
\end{array}
\right. } is valid. Eq. \re{denk_app13} also reduces the number of the
coefficients which has to be calculated, approximately half of the whole set
of the coefficients.

\end{document}